\documentclass[twocolumn,showpacs,preprintnumbers,amsmath,amssymb]{revtex4}
\usepackage{graphicx}% Include figure files
\usepackage{dcolumn}% Align table columns on decimal point

\usepackage{mathptmx, courier, pifont}
\usepackage[scaled=0.92]{helvet}
\usepackage[T1]{fontenc}
\usepackage{textcomp}

\usepackage{bm}% bold math

\newcommand{\tsub}[1]{_{\mbox{\scriptsize#1}}}

\newcommand{\quarterthin}{\kern 0.0417em}

\begin{document}

\title
{$\bm k$-dependent SU(4) model of high-temperature superconductivity
and its coherent-state solutions}

\author{Yang Sun$^{(1)}$}
%\email{ysun@nd.edu}
\author{Mike Guidry$^{(2)}$}
%\email{guidry@utk.edu}
\author{Cheng-Li Wu$^{(3)}$}
%\email{clwu@phys.cts.nthu.edu.tw}

\affiliation{$^{(1)}$Department of Physics, Shanghai Jiao Tong
 University, Shanghai 200240, People's Republic of China \\
$^{(2)}$Department of Physics and Astronomy, University of
Tennessee, Knoxville, Tennessee 37996, USA \\
$^{(3)}$Department of Physics, Chung-Yuan Christian University,
Chungli, Taiwan 320, ROC}

\date{\today}

\begin{abstract}
We extend the SU(4) model \cite{Gu01,Lawu03,Wu04,Sun06,Sun07} for
high-$T\tsub c$ superconductivity to an SU(4)$_k$ model that permits
explicit momentum ($k$) dependence in predicted observables. We
derive and solve gap equations that depend on $k$, temperature, and
doping from the SU(4)$_k$ coherent states, and show that the new
SU(4)$_k$ model reduces to the original SU(4) model for observables
that do not depend explicitly on momentum. The results of the
SU(4)$_k$ model are relevant for experiments such as ARPES that
detect explicitly k-dependent properties. The present SU(4)$_k$
model describes quantitatively the pseudogap temperature scale and
may explain why the ARPES-measured $T^*$ along the anti-nodal
direction is larger than other measurements that do not resolve
momentum. It also provides an immediate microscopic explanation for
Fermi arcs observed in the pseudogap region. In addition, the model
leads to a prediction that even in the underdoped regime, there
exist doping-dependent windows around nodal points in the $k$-space,
where antiferromagnetism may be completely suppressed for all doping
fractions, permitting pure superconducting states to exist.
\end{abstract}


\maketitle

\section{Introduction}

The mechanism that leads to high-temperature superconductivity in
the cuprates remains an open question despite intense study for the
past two decades.  Although the field has been challenged by many
high-quality data from different types of measurement, there is no
uniformly-accepted theoretical picture that can offer a unified and
consistent description for these data. It is often believed that
underlying physics can be understood in terms of individual
particles interacting through appropriately-chosen interactions.
However, even with greatly-simplified Hamiltonians, describing
collective motion in these strongly-correlated, many-electron
systems has had only limited success. This has led some authors
\cite{Anderson} to conclude that the many-body correlations in
cuprates are so strong that  dynamics may no longer be described
meaningfully in terms of electrons and must be described instead in
terms of new effective building blocks that fractionalize spin and
charge.

Simplicity is a kind of beauty in physics.  Even if one could solve
the problem with the help of large-scale numerical calculations,
such a practice may not be interesting because physics can be
completely buried in the numerous configurations used in the
calculation.  On the other hand, alternative approaches to many-body
problems have been proposed. One is the method of fermion dynamical
symmetries \cite{FDSM}. This approach is based on the fact that
collective motions in strongly correlated many-body systems are
often governed by only a few collective degrees of freedom, and a
quantum system exhibiting dynamical symmetries usually contains two
or more competing collective modes.  Once these degrees of freedom
are identified and properly incorporated into a model, the problem
may be considerably simplified and, most importantly, the physics in
such approaches may become transparent.

This is the philosophy that the SU(4) model of high-temperature
superconductivity \cite{Gu01,Lawu03,Wu04,Sun06,Sun07} is based on.
For cuprate systems we have proposed that the most relevant
collective degrees of freedom are $d$-wave superconductivity (SC)
and antiferromagnetism (AF), and that coherent pairs (not individual
particles), formed from two electrons (or holes) centered on
adjacent lattice sites, are appropriate dynamical building blocks of
the wave function.  The choice of this space, which is small in size
but rich in physics, corresponds to a physically-motivated
truncation of the huge Hilbert space corresponding to the original
problem.

It has been found \cite{Gu01} that these spin-singlet ($D$) and the
spin-triplet ($\pi$) pair operators, when supplemented with the
particle-hole type operators for staggered magnetization ($Q$), spin
($S$), and charge, constitute a 16-element operator set that is
closed under a U(4) $\supset$ U(1) $\times$ SU(4) algebra if the
$d$-wave formfactor $g(\bm k)$ in $D$ and $\pi$ pair operators is
replaced by sgn$[g(\bm k)]$. The U(1) factor corresponds to a charge
density wave that is independent of the SU(4) subspace because of
the direct product.  This implies that in the minimal U(4) model
charge density waves do not influence the AF--SC competition in
lowest order and our discussions have been focused in the SU(4)
subspace with its coherent-state approximation \cite{Gu01,Lawu03}.
It has been further discovered \cite{Wu04} that the SU(4) symmetry
is a consequence of non-double-occupancy---the constraint that each
lattice site cannot have more than one valence electron. This
suggests a fundamental relationship between SU(4) symmetry and
Mott-insulator normal states at half-filling for cuprate
superconductors.

Thus, the SU(4) model has ingredients of competing AF and SC modes,
$d$-wave $D$ and $\pi$ pairs (entering as ``preformed pairs" that
are mixtures of the two kinds of paired states under the SU(4)
constraint \cite{Sun06}) in wave functions, and non-double-occupancy
imposed by the symmetry \cite{Wu04}. All of these appear to be
relevant for the physics of cuprate superconductors. For data that
do not resolve an explicit $k$ dependence, the coherent-state
solutions of the SU(4) model (with properly adjusted parameters for
effective interaction strengths) can consistently describe SC gaps,
pseudogaps (PG), and the corresponding transition temperatures
$T\tsub c$ and $T^*$ in cuprates, as demonstrated in Ref.
\cite{Sun07}.

However, there are experimental indications for explicit
$k$-dependence that are observed by experiments such as
angle-resolved photoemission spectroscopy (ARPES)
\cite{Norman98,Kanigel06}. These data probe electrons near the Fermi
surface having particular $k$ directions. To describe $k$-dependence
in energy gaps, we must extend our original SU(4) model by
displaying explicit $k$-dependence in the gap equations and their
corresponding solutions. This is the goal of the present paper.

The paper is organized as follows. In Sec.\ II, we outline the SU(4)
background by pointing out the assumptions made when the original
$k$-independent SU(4) model was constructed. Sections III and IV are
respectively devoted to presentation of the new $k$-dependent
SU(4)$_k$ model and the $k$-dependent gap equations obtained using
the generalized coherent-state method. We solve these gap equations
in Sec.\ V and give analytical solutions for the superconducting gap
and the pseudogap. Finally, we discuss some immediate consequences
of the SU(4)$_k$ model in Sec.\ VI, and a short summary is given in
Sec.\ VII.

\section{Dynamical symmetries and the original SU(4) model}

Interactions in dynamical symmetry theories are determined by
symmetry groups \cite{Gu01}. A general SU(4) Hamiltonian with
pairing and AF interactions can be written as \cite{spinNote}
\begin{equation}
H = H_0 - V_d - V_\pi - V_q,
\label{H}
\end{equation}
where $H_0$ is the single particle (s.p.) energy, and $V_d$,
$V_\pi$, and $V_q$ are the two-body spin-singlet pairing,
spin-triplet pairing, and AF interactions, respectively:
\begin{subequations}
\label{interaction}
\begin{eqnarray}
H_0 &=& \sum_{\bm k} \varepsilon_{\bm k} n_{\bm k}
\label{inter:h0} \\
V_d &=& \sum_{\bm k,\bm k'} G^0_{\bm k \bm k'} D^\dag({\bm
k})D({\bm k'})
\label{inter:vd} \\
V_\pi &=& \sum_{\bm k,\bm k'} G^1_{\bm k \bm k'} \vec
\pi^\dag({\bm k})\cdot\vec \pi({\bm k'})
\label{inter:vpi} \\
V_q &=& \sum_{\bm k,\bm k'} \chi^0_{\bm k \bm k'} \vec Q({\bm
k})\cdot \vec Q({\bm k'}). \label{inte:vq}
\end{eqnarray}
\end{subequations}
The operators appearing in Eqs.\ (\ref{interaction}) can be
expressed as
\begin{subequations}
\label{operator}
\begin{eqnarray}
D^\dag(\bm k) &=& g(\bm k) c_{\bm k\uparrow}^\dag c_{- \bm
k\downarrow}^\dag
\label{Dk}\\
\pi^\dag_{ij}(\bm k) &=& g(\bm k) c_{\bm k+\bm q,i}^\dag c_{\bm
k,j}^\dag
\label{pik}\\
Q_{ij}(\bm k) &=& c_{\bm k+\bm q,i}^\dag c_{\bm k,j}, \label{Qk}
\end{eqnarray}
\end{subequations}
where $\pi^\dag_{ij}(\bm k)$ and $Q_{ij}(\bm k)$ are, respectively,
tensor forms of $\vec \pi^\dag({\bm k})$ and $\vec Q({\bm k})$.  In
Eqs.\ (\ref{operator}), $c_{\bm k,i}^\dagger$ creates an electron of
momentum $\bm k$ and spin projection $i,j= 1 {\rm\ or\ }2 \ (\equiv$
$\uparrow$ or $\downarrow)$, and $\bm q = (\pi,\pi,\pi)$ is an AF
ordering vector. The $d$-wave formfactor,
\begin{equation}
g(\bm k) = g(k_x,k_y) = \cos k_x-\cos k_y, \label{gk}
\end{equation}
appears in (\ref{Dk}) and (\ref{pik}) because of strong
experimental evidence that in cuprates the coherent pairs exhibit
$d$-wave orbital symmetry \cite{Sc95}. Energy gaps thus generally
are $k$-dependent
\begin{subequations}
\label{gap}
\begin{eqnarray}
\Delta_d(\bm k) &=& \sum_{\bm k'} G^0_{\bm k\bm k'} \left< D^\dag(\bm k)\right>\\
\Delta_\pi(\bm k) &=& \sum_{\bm k'} G^1_{\bm k\bm k'} \left< \pi^\dag_z(\bm k')\right>\\
\Delta_q(\bm k) &=& \sum_{\bm k'} \chi^0_{\bm k\bm k'} \left<
Q_z(\bm k')\right>.
\end{eqnarray}
\end{subequations}

The discussion to this point is general and no approximations have
been made. In our original $k$-independent SU(4) model
\cite{Gu01,Lawu03,Wu04,Sun06,Sun07}, we have introduced
approximations through the following assumptions
\begin{subequations}
\label{approximation}
\begin{eqnarray}
g(\bm k) & \approx & \mbox{sgn} [g(\bm k)] \label{gk1}\\
\varepsilon_{\bm k} & \approx & \varepsilon \label{ek}\\
G^i_{\bm k\bm k'} & \approx & G^i \quad (i=0,1), \qquad \chi^0_{\bm
k\bm k'} \approx \chi^0 \label{ggc}.
\end{eqnarray}
\end{subequations}
Assumption (\ref{gk1}) removes the $k$-dependence from formfactors
in the pair operators, and assumptions (\ref{ek}) and (\ref{ggc}),
respectively, replace the s.p.\ energy and interaction strengths
with $k$-independent constants. These approximations thus lead to
$k$-independent gaps
\begin{equation}
\Delta_d = G^0 \left< D^\dag \right> \quad
\Delta_\pi = G^1 \left< \pi^\dag_z \right> \quad
\Delta_q = \chi^0 \left< Q_z \right>,
\label{gap1}
\end{equation}
which are expressed in terms of the collective operators
\begin{subequations}
\label{operator1}
\begin{equation}
\begin{array}{ll}
\displaystyle D^\dag = \sum_{\bm k} \mbox{sgn} [g(\bm k)] c_{\bm
k\uparrow}^\dag c_{-\bm k\downarrow}^\dag \nonumber
\\
\displaystyle \pi^\dag_{ij} = \sum_{\bm k} \mbox{sgn} [g(\bm k)]
c_{\bm k+\bm q,i}^\dag c_{\bm k,j}^\dag \nonumber
\\
\displaystyle Q_{ij} = \sum_{\bm k} c_{\bm k+\bm q,i}^\dag c_{\bm
k,j}. \nonumber
\end{array}
\end{equation}
\end{subequations}
The preceding equations constitute the basis for discussions in
Refs.\ \cite{Gu01,Lawu03,Wu04,Sun06,Sun07}, and all our previous
SU(4) results are obtained within this framework. As most cuprate
data presumably represent weighted averages over contributions of
different $k$ components, the original SU(4) scheme works well. In
Ref.\ \cite{Sun06}, we derived and solved $k$-independent (but
temperature and hole-doping dependent) SU(4) gap equations, and
used the results to construct generic gap and phase diagrams. We
compared the results with some representative cuprate data in
Ref.\ \cite{Sun07} and found that, for data that do not resolve an
explicit $k$ dependence, the coherent-state solutions of the
original SU(4) model can consistently describe SC gaps,
pseudogaps, and the corresponding transition temperatures $T\tsub
c$ and $T^*$.

\section{The $\bm k$-dependent SU(4)$_k$ model}

As we have noted, there is experimental evidence for explicit
$k$-dependence of certain observables in the cuprates. Although
interpretation of some of the results remains somewhat
controversial, their momentum-dependent nature is clear. One example
is the observation of Fermi arcs in angle-resolved photoemission
spectroscopy (ARPES) data \cite{Norman98,Kanigel06}: ARPES
measurements suggest that the Fermi surface is gapped out arcwise in
the pseudogap region below $T^*$, indicating clear anisotropy of the
PG in the $k$-space.

In order to describe momentum dependence of energy gaps, we must
extend our original SU(4) model \cite{Gu01,Lawu03,Wu04,Sun06,Sun07}
in a way that restores the $k$-dependence that is washed out by the
assumptions in Eqs.\ (\ref{approximation}), but preserves the SU(4)
symmetry. The replacement (\ref{gk1}) for the pair operators is a
necessary condition for preserving the SU(4) algebra \cite{Gu01},
which is required physically because it imposes the
non-double-occupancy condition \cite{Wu04}. Therefore, the only way
to restore $k$-dependence in energy gaps but keep the SU(4) symmetry
(and its associated non-double-occupancy constraint) is to modify
(\ref{ek}) and (\ref{ggc}) to allow the s.p.\  energy $\varepsilon$
and the interaction strengths to carry $k$-dependence. The s.p.\
energy term $\varepsilon$ is less important in this regard because
it does not contribute to energy gaps and transition temperature in
our formalism \cite{Sun06}. We may thus employ it in the most
general form $\varepsilon_{\bm k}$.

\begin{figure}
  \includegraphics[height=.20\textheight]{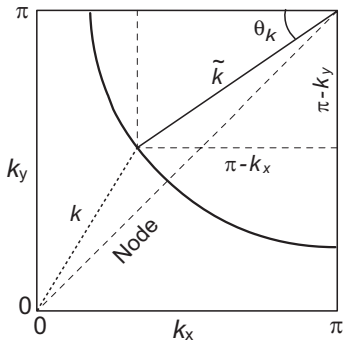}
  \caption{Geometry and definitions in the $k_x$--$k_y$ plane,
  where $\tilde k$ is hole momentum.}
\end{figure}

\begin{figure}
  \includegraphics[height=.20\textheight]{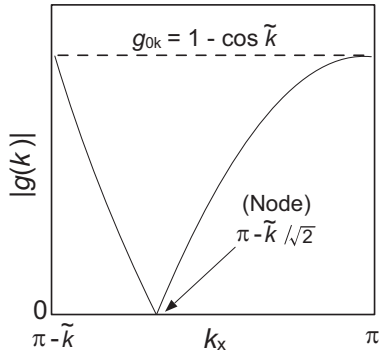}
  \caption{The curve $|g(k)|$ and its maximum value $g_{0k}=1-\cos \tilde k$
   for a given momentum $\bm k=(k_x,k_y)$, under the constraint (\ref{ktilde}).}
\end{figure}

Without loss of generality, $g(\bm k)$ can be written as the
product of the absolute value and a sign
\begin{equation}
g(\bm k)=|g(\bm k)| \times \mbox{sgn}[g(\bm k)]. \label{gk2}
\end{equation}
Therefore, the approximation (\ref{gk1}) implies that in our
original SU(4) model we have assumed that the magnitude $|g(\bm k)|$
is unity, regardless of $\bm k$. (Note that this is also the
condition used to close the algebra of the SO(5) model
\cite{Kohno98,Hen98}.) Instead of $|g(\bm k)|=1$, we now introduce
\begin{equation}
|g(\bm k)| \approx \sigma_{\bm k} = g_{0k} \delta(\theta_k),
\label{gk3}
\end{equation}
where $g_{0k}$ is the maximum value of $|g(\bm k)|$. In Figs.~1
and 2, we illustrate the associated geometry and definitions. In
our notation, $\bm k = (k_x,k_y)$ is the electron momentum under
the constraint
\begin{equation}
(\pi-k_x)^2 + (\pi-k_y)^2 =
\tilde k^2 ,
\label{ktilde}
\end{equation}
where $\tilde k$ is the hole momentum with $\theta_k$ its azimuthal
angle, as shown in Fig.~1. In Eq.~(\ref{gk3}), $\delta(\theta_k)$
takes a value of unity except in a narrow region around the nodal
points (corresponding to $|k_x|=|k_y|$ or $\theta_k = \pm\pi/4$ for
the first Brillouin zone; see Fig.~1), where it quickly diminishes
and vanishes exactly at the nodal points. A possible mathematical
expression could be of the Gaussian type
\begin{equation}
\delta(\theta_k)=1-e^{-\left({{\theta_k -\pi/4}\over
{\Delta\theta}}\right)^2} ~~~{\rm with}~~~ \Delta\theta\ll
\pi/4,\nonumber
\end{equation}
where $\Delta\theta$ measures width of the Gaussian. With very small
$\Delta\theta$, the exponential term has a negligible contribution
to the average, which ensures the averaged $\delta(\theta_k)$ equal
to 1. The so-defined $\delta(\theta_k)$ becomes exactly zero at
$\theta_k = \pm\pi/4$. Therefore, our pairing gaps have nodes at
$\theta_k = \pm\pi/4$, which agrees with experiments.

The behavior of $|g(\bm k)|$ is illustrated in Fig.\ 2. It is easy
to show that
\begin{equation}
g_{0k}=\left| 1-\cos \tilde k \right|.
\label{gzerok}
\end{equation}
Thus, for the first Brillouin zone $k_x$ and $k_y$ can take values
from zero to $\pi$, while $g_{0k}$ changes from 0 to 2. (The
assumption $|g(\bm k)|=1$ in the original SU(4) model is thus
equivalent to taking an average of $g_{0k}$ over $\bm k$.) Equation
(\ref{gk3}), with its explicit dependence on $\bm k$, clearly
improves on the original SU(4) model for observables having a
possible $k$ dependence.

With the new approximation (\ref{gk3}) for $|g(\bm k)|$, the
pairing interaction strengths in Eq.\ (\ref{interaction}) are
\begin{equation}
G^i_{\bm k\bm k'} = G^0_i\sigma_{\bm k}\sigma_{\bm k'} \quad
(i=0,1). \label{newg}
\end{equation}
The factor $\mbox{sgn}[g(\bm k)]$ in the product (\ref{gk2}) remains
unchanged in the pair operators, which ensures preservation of the
SU(4) symmetry.

The $k$-dependence of the AF interaction $\chi_{\bm k\bm k'}$
follows from the nature of exchange interactions. The corresponding
matrix elements are proportional to the wavefunction overlap between
the states, which are one-particle, one-hole states with momenta
$(\bm k+\bm q, \bm k)$. The $d$-wave symmetry in the pair structure
implies that the amplitude of a pair wavefunction with two electrons
having momenta $(\bm k, -\bm k)$ in the background mean-field of the
SU(4) collective subspace is proportional to $g(\bm k)$.  This means
physically that the two electrons in a pair favor aligning their
momenta $\bm k$ along the Cu--O bond direction (maximum of $|g(\bm
k)|$) rather than along the diagonal to the Cu--O bonds (nodal
direction of $g(\bm k)$; see Fig.\ 1). Since the wavefunction for
momentum $\bm k+\bm q$ differs from that for $\bm k$ only by a sign
\cite{Gu01}, and a hole wavefunction is the conjugate of a particle
wavefunction, the amplitude of a particle--hole wavefunction with
momenta $(\bm k+\bm q, \bm k)$ should be very similar to that of a
pair and thus proportional to $g(\bm k)$. Therefore, we can write
\begin{equation}
\chi_{\bm k\bm k'} =  \chi^0 g_{\bm k} g_{\bm k'}, \label{newchi}
\end{equation}
where $g_{\bm k}\equiv |g(\bm k)|$. Note that for the $g_{\bm k}$
factor in Eq.\ (\ref{newchi}), no approximation like the one in Eq.\
(\ref{gk3}) is necessary. Explicitly, we mean here that $g_{\bm k} =
|g(\bm k)| = |\cos k_x -\cos k_y|$.

Inserting Eqs.\ (\ref{newg}) and (\ref{newchi}) into
(\ref{interaction}), we can rewrite the Hamiltonian (\ref{H}) as
\begin{eqnarray}
H &=& \sum_{\bm k} \varepsilon_{\bm k} n_{\bm k} - G^0_0 \sum_{\bm
k,\bm k'} \sigma_{\bm k}\sigma_{\bm k'} D^\dag(\bm k)D(\bm k')
\nonumber\\
&&- G^0_1 \sum_{\bm k,\bm k'} \sigma_{\bm k}\sigma_{\bm k'} \vec
\pi^\dag(\bm k)\cdot\vec \pi(\bm k')
\nonumber\\
&&- \chi^0 \sum_{\bm k,\bm k'} g_{\bm k} g_{\bm k'} \vec Q(\bm
k)\cdot \vec Q(\bm k'). \label{Hk}
\end{eqnarray}
The $k$-dependent Hamiltonian (\ref{Hk}) possesses a $\sum_k
\otimes$ SU(4)$_k$ symmetry, with 15 $k$-dependent generators
\begin{eqnarray}
{\bf D}^\dag (\bm k) &=& D^\dag (\bm k) + D^\dag (-\bm k) \nonumber\\
\vec{\bm \pi}^\dag (\bm k) &=& \vec\pi^\dag (\bm k) + \vec\pi^\dag (-\bm k) \nonumber\\
\vec {\bf Q} (\bm k) &=& \vec Q (\bm k) + \vec Q (-\bm k) \nonumber\\
     {\bf M} (\bm k) &=& M (\bm k) + M (-\bm k) \nonumber\\
\vec {\bf S} (\bm k) &=& \vec S (\bm k) + \vec S (-\bm k),
\nonumber
\end{eqnarray}
where $\bf M (\bm k)$ and $\vec {\bf S} (\bm k)$ are, respectively,
the charge and the spin operators. For each $\bm k$, the commutation
relation among generators, the structure of subgroup chains, and
their corresponding properties are analogous to those of the
original SU(4) group structure \cite{Gu01}. We term this
$k$-dependent extension \cite{Dukelsky07} of the original SU(4)
model the SU(4)$_k$ model.

\section{$\bm k$-dependent gap equations}

In the preceding section, we demonstrated that with a better
approximation to the absolute value of the formfactor $g(\bm k)$ it
is possible to introduce explicit $k$ dependence through a symmetry
structure that corresponds to a product of SU(4) groups, each
labeled by $k$. Therefore, the following discussions for gap
equations and their solutions for a given $k$ are rather similar to
those in Ref.\ \cite{Sun06} for the $k$-independent SU(4) model.

By analogy with discussions in Appendix B of Ref.\ \cite{Sun06},
under the coherent-state (symmetry-constrained, generalized
Hartree--Fock--Bogoliubov) approximation, one obtains for the
$k$-dependent case
\begin{equation}
2u_{\bm k\pm}v_{\bm k\pm} (\varepsilon_{\bm k\pm}-\lambda) -
\Delta_{\bm k\pm} (u^2_{\bm k\pm} - v^2_{\bm k\pm}) = 0
\label{BCSext}
\end{equation}
with
$$
\varepsilon_{\bm k\pm} = \varepsilon_{\bm k} \mp \Delta_q(\bm k)
\qquad \Delta_{\bm k\pm} = \Delta_d(\bm k) \pm \Delta_\pi(\bm k)
\nonumber
$$
and
\begin{eqnarray}
\Delta_d(\bm k) &=& G^0_0 \sigma_{\bm k} \sum_{\bm k'>0} \sigma_{\bm
k'} \left< {\bf D}^\dag(\bm k')\right> \nonumber\\
\Delta_\pi(\bm k) &=& G^0_1 \sigma_{\bm k} \sum_{\bm k'>0}
\sigma_{\bm k'} \left< {\bm \pi}_z^\dag(\bm k')\right> \nonumber\\
\Delta_q(\bm k) &=& \chi^0 g_{\bm k} \sum_{\bm k'>0} g_{\bm k'}
\left< {\bf Q}_z (\bm k')\right>. \nonumber
\end{eqnarray}
${\bm k'>0}$ in the above and following equations means $k'_x
>0$ or $k'_y >0$. Solving Eq.\ (\ref{BCSext}) gives $k$-dependent
occupation probabilities
$$
u^2_{\bm k\pm} = \frac12  \left[ 1 + {{\varepsilon_{\bm
k\pm}-\lambda}\over {e_{\bm k\pm}}}\right] \qquad v^2_{\bm k\pm} =
\frac12 \left[ 1 - {{\varepsilon_{\bm k\pm}-\lambda}\over {e_{\bm
k\pm}}}\right]
$$
and a quasiparticle energy
\begin{equation}
e_{\bm k\pm} = \sqrt{(\varepsilon_{\bm k\pm}-\lambda)^2 +
\Delta_{\bm k\pm}^2}. \nonumber
\end{equation}
The gap equations in $k$-space can then be obtained:
\begin{subequations}
\label{kgapeqs}
\begin{eqnarray}
\Delta_d(\bm k) &=& {{G^0_0\sigma_{\bm k}}\over 2} \sum_{\bm k'
> 0} \sigma_{\bm k'} \left( w_{\bm k'+}\Delta_{\bm k'+}
+ w_{\bm k'-}\Delta_{\bm k'-} \right) \\
\Delta_\pi(\bm k) &=& {{G^0_1\sigma_{\bm k}}\over 2} \sum_{\bm k'
> 0}\sigma_{\bm k'}\left( w_{\bm k'+}\Delta_{\bm k'+}
- w_{\bm k'-}\Delta_{\bm k'-} \right) \\
\Delta_q(\bm k) &=& {{\chi^0 g_{\bm k}}\over 2} \sum_{\bm k' > 0}
g_{\bm k'}
 \{ w_{\bm k'+} [\Delta_q(\bm k') + \lambda'_{\bm k'}] \nonumber\\
&&+ w_{\bm k'-} [\Delta_q(\bm k') - \lambda'_{\bm k'}] \} \\
-2x &=& {2\over \Omega} \sum_{\bm k' > 0} \{ w_{\bm k'+}
[\Delta_q(\bm k') + \lambda'_{\bm k'}] \nonumber\\
&&- w_{\bm k'-} [\Delta_q(\bm k') - \lambda'_{\bm k'} ] \}
\label{15d}
\end{eqnarray}
\end{subequations}
with
\begin{equation}
\begin{array}{c}
w_{\bm k\pm} = \displaystyle {{P_{\bm k\pm}(T)}\over {e_{\bm k\pm}}}
\qquad \lambda'_{\bm k} = \lambda - \varepsilon_{\bm k}
\\[5pt]
P_{\bm k\pm}(T) = \displaystyle \mbox{tanh}\left({{e_{\bm
k\pm}}\over{2k\tsub BT}}\right). \nonumber
\end{array}
\end{equation}
In Eq.\ (\ref{15d}), $\Omega=\sum_{\bm k>0}$ is the maximum number
of doped holes (or doped electrons for electron-doped compounds)
that can form coherent pairs, assuming the normal state (at half
filling) to be the vacuum. $x$ is the relative doping fraction in
the model \cite{Sun06}. Positive $x$ represents the case of hole
doping, with $x=0$ corresponding to half filling (no doping) and
$x=1$ to maximal hole doping. The true doping $P$ is related to $x$
by $x \simeq 4P$ \cite{Sun06}.

The $k$-dependent gap equations (\ref{kgapeqs}) are coupled
algebraic equations. By solving these equations, one can in
principle obtain $k$-dependent (also temperature and hole-doping
dependent) energy gaps. However, general and exact solutions are
difficult because gaps for given $\bm k$ are related to all other
$\bm k$ points, which means that solutions at each $\bm k$ point are
not independent from the other $k$-components. In the next section,
we show that one can obtain analytical solutions by applying some
approximations.

\section{Solutions for $\bm k$-dependent gap equations}

We may greatly simplify the solution of Eqs.\ (\ref{kgapeqs})
through the following three steps. First, we replace the quantities
in the summations on the right hand side of (\ref{kgapeqs}) with
their corresponding mean values:
\begin{equation}
\begin{array}{c}
\Delta_{\bm k\tau} \Longrightarrow \overline{\Delta}_\tau
  (\tau=+, - ,q) \qquad
\lambda'_{\bm k} \Longrightarrow \overline{\lambda'}
\\[5pt]
w_{\bm k\pm} \Longrightarrow \overline{w}_\pm \equiv \displaystyle
{{\overline P_\pm(T)}\over {\overline e_\pm}}, \nonumber
\end{array}
\end{equation}
with
\begin{equation}
\overline P_{\pm}(T) = \mbox{tanh}\left({{\overline
e_{\pm}}\over{2k\tsub BT}}\right) \qquad \overline e_{\pm} =
\sqrt{(\overline {\lambda'}\pm \overline \Delta_q)^2 + \overline
\Delta_\pm^2}. \nonumber
\end{equation}
The functions $\sigma_{\bm k}$ and $g_{\bm k}$ in the summations can
then be simplified as
\begin{eqnarray}
\sum_{\bm k'>0} \sigma_{\bm k'} &\Longrightarrow&
\sum_{\bm k'>0} \overline g_0 = {\Omega\over 2} \overline g_0 \label{intg0}\\
\sum_{\bm k'>0} g_{\bm k'} &\Longrightarrow& \sum_{\bm k'>0}
\overline g = {\Omega\over 2}\overline g. \label{intg}
\end{eqnarray}

The second level of simplification is based on physical
considerations. Experimentally-measured energy gaps are dominated by
contributions from near the Fermi surface. Therefore, we assume that
measured gaps may be approximated by their values at $\tilde
k=k\tsub f$. Using this approximation and the average values
introduced in the first approximation step, we can write for the gap
equations of (\ref{kgapeqs}) evaluated at $\tilde k=k\tsub f$:
\begin{subequations}
\label{kgapseqs2}
\begin{eqnarray}
\Delta_d(\bm k) &=& {\Omega\over 4}  G^0_0 g_0 \overline g_0
\delta(\theta_{k\tsub f})  \left( \overline w_{+}\overline\Delta_{+}
+ \overline w_{-} \overline\Delta_{-} \right) \\
\Delta_\pi(\bm k) &=& {\Omega\over 4}  G^0_1 g_0 \overline g_0
\delta(\theta_{k\tsub f}) \left( \overline w_{+} \overline\Delta_{+}
- \overline w_{-} \overline\Delta_{-} \right) \\
\Delta_q(\bm k) &=& {\Omega\over 4}  \chi^0 g_0 \overline g
\gamma(\theta_{k\tsub f}) \{ \overline w_{+} [\overline\Delta_q +
\overline{\lambda'}]
\nonumber\\
&&+ \overline w_{-} [\overline\Delta_q - \overline{\lambda'} ]\} \\
-2x &=& \overline w_{+} [\overline\Delta_q + \overline{\lambda'}] -
\overline w_{-} [\overline\Delta_q - \overline{\lambda'} ] ,
\end{eqnarray}
\end{subequations}
where $g_0 \equiv g_{0k\tsub f}$, and
\begin{equation}
\gamma(\theta_{k\tsub f}) \equiv  \left| {\frac{g(\bm k)}{g_0}}
\right|. \label{gammaOfTheta}
\end{equation}
Equations (\ref{kgapseqs2}) are $k$-dependent gap equations
constrained on the Fermi surface through
\begin{equation}
(\pi-k_x)^2 + (\pi-k_y)^2 = k\tsub f^2. \label{fSurface}
\end{equation}
It can be shown that $\gamma(\theta_{k\tsub f})$ is independent of
$|k\tsub f|$, to high accuracy, and therefore can be considered in
later discussions to be a function of azimuthal angle only.

In the third simplification step, we assume the average values
$\overline\Delta_\tau$ and $\overline{\lambda'}$ to be proportional
to the unknown quantities $\Delta_{\tau}(\bm k)$ and $\lambda'_{\bm
k}$, respectively, with a constant of proportionality $R$:
\begin{equation}
\overline \Delta_\tau = R \Delta_{\tau}(\bm k) \quad (\tau=+, -, q)
\qquad \overline {\lambda'} = R \lambda'_{\bm k}, \label{assum3}
\end{equation}
which implies that
\begin{equation}
\overline e_\pm = R e_{\bm k \pm}.
\end{equation}
The parameter $R$ serves as a renormalization factor that corrects
on average for the errors caused by the approximation and is
determined by fitting data. With (\ref{assum3}), Eqs.\
(\ref{kgapseqs2}) now become
\begin{subequations}
\label{kgapseqs3}
\begin{eqnarray}
\Delta_d(\bm k) &=& {\Omega\over 4} G_{0} \delta(\theta_{k\tsub f})
\left[ \tilde w_{+} \Delta_{+}(\bm k) + \tilde w_{-} \Delta_{-}(\bm
k) \right]
\\
\Delta_\pi(\bm k) &=&  {\Omega\over 4} G_{1} \delta(\theta_{k\tsub
f}) \left[ \tilde w_{+} \Delta_{+}(\bm k) - \tilde w_{-}
\Delta_{-}(\bm k) \right]
\\
\Delta_q(\bm k) &=& {\Omega\over 4} \chi {{\gamma(\theta_{k\tsub
f})}\over{\overline\gamma}}
\{ \tilde w_{+} [\Delta_{q}(\bm k) + \lambda'_{\bm k}] \nonumber\\
&&+ \tilde w_{-} [\Delta_{q}(\bm k) - \lambda'_{\bm k} ]\}
\\
-2x &=& \tilde w_{+} [\Delta_{q}(\bm k) + \lambda'_{\bm k}]
\nonumber\\
&& -\tilde w_{-} [\Delta_{q}(\bm k) - \lambda'_{\bm k} ],
\end{eqnarray}
\end{subequations}
with
\begin{equation}
G_{i} = G^0_i g_0 \overline g_0 \qquad  \chi = \chi^0 g_0
\overline\gamma \overline g \nonumber
\end{equation}
and
\begin{equation}
\tilde w_{\pm} = {{\tilde P_{\pm}(T)}\over {e_{\bm k\pm}}} \qquad
\tilde P_{\pm}(T) = \mbox{tanh}\left({{R e_{\bm k\pm}}\over{2k\tsub
BT}}\right).
\end{equation}
In the above equations, $\overline\gamma$  is the average value of
$\gamma(\theta_{k\tsub f})$.

The simplified gap equations (\ref{kgapseqs3}) can now be solved
analytically. They have the same structure as the gap equations
discussed in the $k$-independent SU(4) model \cite{Sun06}, except
that the interaction strengths in the present case are
$k$-anisotropic. Therefore, all the SU(4) formulas in Sections
III--VI of Ref.\ \cite{Sun06} remain valid, provided that the
following replacements are made for the singlet-pairing,
triplet-pairing, and antiferromagnetic coupling strengths,
respectively:
\begin{equation}
G_0 \rightarrow G_0 \delta(\theta_{k\tsub f}) \quad\ G_1 \rightarrow
G_1 \delta(\theta_{k\tsub f}) \quad\ \chi \rightarrow \chi
\gamma(\theta_{k\tsub f}) / \overline\gamma.
\end{equation}
For example, if we introduce the the doping parameter $x$ defined in
\S II.c of Ref.\ \cite{Sun06}, the $k$-dependent critical
hole-doping fraction is [compare Eq.\ (23) of Ref.\ \cite{Sun06}]
\begin{equation}
x_{q\theta} \equiv x_q(\theta_{k\tsub f}) = \sqrt{{\chi
\gamma(\theta_{k\tsub f}) / \overline\gamma -
G_0\delta(\theta_{k\tsub f})} \over {\chi \gamma(\theta_{k\tsub
f})  / \overline\gamma - G_1\delta(\theta_{k\tsub f})}},
\label{xqf}
\end{equation}
and the $T=0$ energy gaps at the Fermi momentum $k\tsub f$ for $x\le
x_{q\theta}$ are obtained as
\begin{subequations}
\label{solution1}
\begin{eqnarray}
\Delta_d(\bm k) &=& {\Omega\over 2} G_0\delta(\theta_{k\tsub f})
 \sqrt{x(x_{q\theta}^{-1}-x)} \\
\Delta_\pi(\bm k) &=& {\Omega\over 2} G_1\delta(\theta_{k\tsub f})
 \sqrt{x(x_{q\theta}-x)} \\
\Delta_q(\bm k) &=& {\Omega\over 2} \chi {{\gamma(\theta_{k\tsub
f})}\over {\overline\gamma}}
 \sqrt{(x_{q\theta}^{-1}-x)(x_{q\theta}-x)} \\
\lambda'_{\bm k} &=& - {\Omega\over 2} \left[\chi
{{\gamma(\theta_{k\tsub f})}\over {\overline\gamma}} -
G_1\delta(\theta_{k\tsub f})\right] x_{q\theta}
\left(1-x_{q\theta}x\right)
\nonumber\\
&&  - {\Omega\over 2} G_1\delta(\theta_{k\tsub f})  x,
\end{eqnarray}
\end{subequations}
while for $x>x_{q\theta}$ we obtain $\Delta_{q}(\bm k) =
\Delta_\pi(\bm k)=0$ and
\begin{subequations}
\label{solution2}
\begin{eqnarray}
\Delta_d(\bm k) &=& {\Omega\over 2} G_0\delta(\theta_{k\tsub f})  \sqrt{1-x^2} \\
\lambda'_{\bm k} &=& -{\Omega\over 2} G_0\delta(\theta_{k\tsub f}) x
\end{eqnarray}
\end{subequations}
for the solutions. The energy gaps obtained in Eqs.\
(\ref{solution1})--(\ref{solution2}) are $k$-anisotropic. The
pairing gaps have nodal points at $k_x=k_y$, where $
\delta(\theta_{k\tsub f})=0. $ The pseudogap $\Delta_{q}(\bm k)$ is
a function of $\bm k$ by virtue of the factor $\gamma(\theta_{k\tsub
f})$ defined in Eq.\ (\ref{gammaOfTheta}).

Because all SU(4) formulas in Sections III--VI of Ref.\ \cite{Sun06}
remain valid, it is easily proven that the PG closure temperature
$T^*$ acquires the same $g(\bm k)$ dependence as the pseudogap, and
we obtain for the PG closure temperature
\begin{equation}
T^*(\bm k) = \chi {{\gamma(\theta_{k\tsub f})}\over {
\overline\gamma}} \Omega {{R(1-x^2)} \over {4k\tsub B}} .
\label{Tstark}
\end{equation}
We do not expect a corresponding effect in the superconducting
region because below $T\tsub c$ the pairing gap opens and the entire
Fermi surface will be destroyed except at the nodal points. We find
a superconducting transition temperature
\begin{equation}
T\tsub c(\bm k) = G_0 \delta(\theta_{k\tsub f}) \Omega {{Rx} \over
{4k\tsub B \, {\mbox {atanh}}(x)}},
\end{equation}
which has no $g(\bm k)$ factor.

\section{Discussions and predictions}

Most experimental techniques do not resolve $k$ and we expect for
those that transition temperatures are dominated by contributions
from near the Fermi surface $(\tilde k = k\tsub f)$, averaged over
all $k$-directions. If one takes the average over $\theta_{k\tsub
f}$, then
$$
\delta(\theta_{k\tsub f})\rightarrow 1
\qquad \gamma(\theta_{k\tsub f})/\overline\gamma\rightarrow 1,
$$
and the gap equations of the $k$-dependent SU(4)$_k$ model and their
solutions become identical to those obtained for the original
$k$-independent SU(4) model \cite{Sun06}. Thus our original SU(4)
model predicts \cite{Sun06,Sun07} values of energy gaps and the
corresponding transition temperatures that are (perhaps weighted)
{\em averages over $k$.} These are relevant for comparison with
experiments that do not resolve $k$. However, the explicit
appearance of the anisotropic factor $\gamma(\theta_{k\tsub
f})/\overline\gamma$ in the gap solutions of the SU(4)$_k$ model
leads to some interesting new consequences. We note that although
our following discussions are made through the anisotropic factor
$\gamma(\theta_{k\tsub f})/\overline\gamma$, its relation with
$\delta(\theta_{k\tsub f})$ guarantees that the model still
preserves the $d$-wave nature and has nodes in the pairing gaps. In
this section we discuss three predictions following from the new
formalism that could have important implications for experiments
that detect explicitly $k$-dependent properties.

\subsection{Two pseudogap closure temperatures:
the maximum and the averaged}

The $k$-dependent PG closure temperature $T^*(\bm k)$ in
Eq.~(\ref{Tstark}) differs from the $k$-averaged one derived in
Eq.~(49) of Ref.~\cite{Sun06},
\begin{equation}
T^*_{\rm ave} = \chi \Omega {{R(1-x^2)} \over {4k\tsub B}} ,
\label{Tstar}
\end{equation}
by the factor ${\gamma(\theta_{k\tsub f})} / {\overline\gamma}$. We
know that ${\gamma(\theta_{k\tsub f})}$, and also $T^*(\bm k)$, take
their maximum values at the antinodal points; for example,
$$
{\gamma_{\rm
max}(\theta_{k\tsub f})}|_{\theta_{k\tsub f}=0,{\pi\over 2}} = 1
$$
(see Figs.~1 and 2). We can denote the maximum PG closure
temperature as $T^*_{\rm max}$. Thus, the SU(4)$_k$ model predicts
two PG closure temperatures that are related to each other through
\begin{equation}
T^*_{\rm max} = T^*_{\rm ave}/ {\overline\gamma} .
\label{TstarRatio}
\end{equation}
It is straightforward to evaluate the averaged $\gamma$ value by
integration. Noting that Eqs. (\ref{solution1}) are restricted to
$x\le x_{q\theta}$, we can define the maximum allowed azimuthal
angle $\theta\tsub c$ through the condition $x=
x_{q\theta}(\theta\tsub c)$. We then have
\begin{equation}
{\overline\gamma} = {2\over\pi} \int_0^{\theta\tsub c}
\left|{{g(k(\theta))}\over {g_0}}\right| d\theta .
\label{gammaBar}
\end{equation}
In the above calculation, we have used for the integrand the
expression (\ref{gammaOfTheta}) for ${\gamma(\theta_{k\tsub f})}$
and the constraint (\ref{fSurface}). The resulting
${\overline\gamma}$ depends on the size of the Fermi surface
$|k\tsub f|$. Assuming an isotropic hole Fermi surface, we have
\begin{equation}
{k\tsub f}^2 = 2\pi (1+P) .\nonumber
%\label{gammaBar}
\end{equation}
Therefore, ${\overline\gamma}$ is essentially a
hole-doping-dependent quantity. In Fig.~3, we show the behavior of
$1 / {\overline\gamma}$ as a function of doping $P$ assuming
coupling-strength parameters characteristic of the cuprate
superconductors. As one can see, it has a nonlinear dependence on
doping, taking the maximum value 1.6 at very small dopings, falling
rapidly between $P=0.05$ and 0.08, and continuously decreases but
with a smaller rate until it reaches unity at the critical doping
$P=0.18$.

\begin{figure}
  \includegraphics[height=.23\textheight]{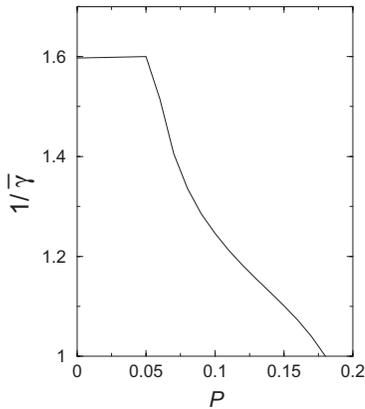}
\caption{The doping-dependent $1 / {\overline\gamma}$ factor. The
calculation employs Eq.~(\ref{gammaBar}) and utilizes realistic
interaction strengths $\chi, G_0$ and $G_1$ taken from
Ref.~\cite{Sun07}, with the pairing onset at $P=0.05$. }
\end{figure}

\begin{figure}
  \includegraphics[height=.27\textheight]{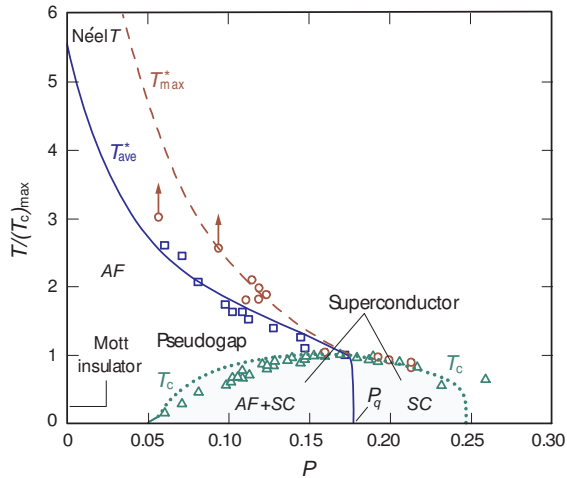}
\caption{(Color online) SU(4) cuprate phase diagram compared with
data. Strengths of the AF and singlet pairing correlations were
determined in Ref. \cite{Sun07} by global fits to cuprate data. The
PG temperature is $T^*$ and the SC transition temperature is $T\tsub
c$. The AF correlations vanish, leaving a pure singlet $d$-wave
condensate, above the critical doping  $P_q$. Dominant correlations
in each region are indicted by italic labels. Data in green (open
triangles) and blue (open squares) are taken from Ref.\
\cite{Dai99}, and those in red (open circles) from Ref.\
\cite{Camp99} (arrows indicate that the point is a lower limit).}
\end{figure}

\begin{figure*}
  \includegraphics[height=.19\textheight]{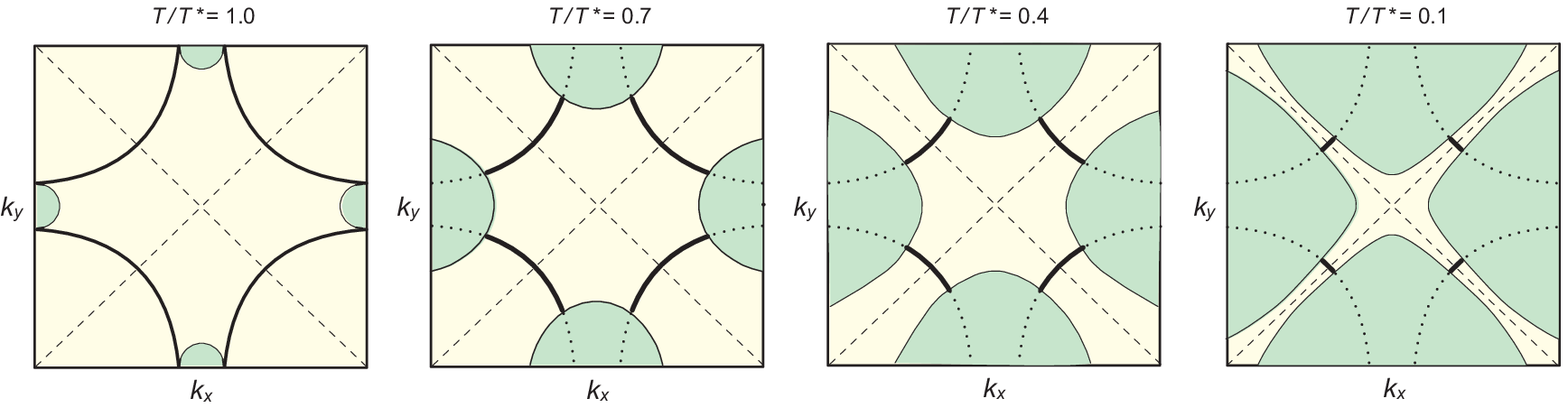}
\caption{(Color online) Construction of Fermi arcs for doping
$P=0.15$ and values of $T/T^*_{\rm max}$ decreasing left to right.
The ranges of $k_x$ and $k_y$ are $-\pi$ to $\pi$ and dashed lines
indicate nodes. Hole Fermi surfaces in the absence of gaps are
illustrated by full solid arcs in each corner. For $T \ge T^*_{\rm
max}$ a full Fermi surface exists; for $T<T^*_{\rm max}$, opening of
the pseudogap destroys Fermi surfaces in the shaded regions (dotted
lines), leaving arcs (solid lines) centered on the nodal lines.
These arcs have absolute lengths that depend on $P$ and $T$, but
relative lengths that depend only very weakly on $P$ and are
determined almost entirely by the ratio $T/T^*_{\rm max}$. The sizes
of the shaded regions grow with decreasing $T$, so at very low
temperature almost all of the Fermi surface becomes gapped and the
Fermi arcs shrink to the nodal points as $T/T^*_{\rm max}
\rightarrow 0$. }
\end{figure*}

We thus obtain two distinct PG closure temperatures, $T^*_{\rm ave}$
and $T^*_{\rm max}$, having the same microscopic origin
\cite{Sun06,Sun07}, but differing  in the kinds of experimental
observables for which they are appropriate. The largest difference
between the two is found for small doping; they take similar values
at large dopings, becoming identical at the optimal doping point. In
Ref.~\cite{Sun07}, experimental values of $T\tsub c$ and $T^*$
\cite{Dai99} that do not resolve $k$ were compared to our
theoretical $T\tsub c$ and $T^*_{\rm ave}$. In Fig.~4, we re-plot
these values in green for $T\tsub c$ (with open triangles for data
and dotted curve for theory) and blue for $T^*_{\rm ave}$ (with open
squares for data and solid curve for theory). Above the blue (solid)
curve, we now add the maximum PG closure temperatures $T^*_{\rm
max}$ in red (with open circles for data and dashed curve for
theory). Because of the $1 / {\overline\gamma}$ factor shown in
Fig.~3, the red (dashed) curve lies well above the blue (solid)
curve at low dopings. In Ref.~\cite{Camp99}, Campuzano {\it et al.}
reported their ARPES data (plotted as red circles in Fig.~4). In the
underdoped regime it is clear that the results from
Refs.~\cite{Camp99} and  \cite{Dai99} differ substantially. Because
the ARPES experiment typically detects $k$-dependent properties, we
suggest that the data from Ref.~\cite{Camp99} actually measure
$T^*_{\rm max}$, as predicted in the present paper, while the data
cited in Ref.~\cite{Dai99} measure the $k$-averaged $T^*_{\rm ave}$,
as described in our earlier paper (Ref.~\cite{Sun07}). We emphasize
that in this interpretation the two types of experiments are seeing
the {\em same underlying physics,} but the observations differ
because what is actually being measured differs in the two cases.

\subsection{Temperature-dependent Fermi arcs}

The pseudogap closure temperature $T^*(\bm k)$ is anisotropic in
$\bm k$. Combining Eqs.\ (\ref{Tstark}), (\ref{Tstar}) and
(\ref{TstarRatio}), we have
\begin{equation}
T^*(\bm k) = T^*_{\rm max} \gamma(\theta_{k\tsub f}) = {{T^*_{\rm
max}}\over {g_0}} |g({k\tsub f})|, \label{fg1.1}
\end{equation}
where the doping-dependent quantity $T^*_{\rm max}$ is the maximum
value of $T^*(\bm k)$ in the antinodal direction, $\theta_{k\tsub
f}=0$ or $\pi/2$. For an arbitrary temperature $T < T^*_{\rm max}$,
the $k$-dependent pseudogap closes when $T=T^*(\bm k)$, which is
equivalent to the requirement that
\begin{equation}
|\cos k_x -\cos k_y| = g_0 (T/T^*_{\rm max}), \label{fg1.2}
\end{equation}
upon substituting (\ref{gk}) and (\ref{fg1.1}). This equation says
that the magnitude of the $d$-wave formfactor (\ref{gk}) that
expresses the nodal structure \cite{Sc95} in cuprate superconductors
is proportional to the scaled quantity $T/T^*_{\rm max}$, with a
proportionality factor $g_0$ that is related to the size of the
Fermi surface.

Simultaneous solution of Eqs. (\ref{fSurface}) and (\ref{fg1.2})
gives values of $k_x$ and $k_y$ where the pseudogap closes at a
given $P$ and $T$. Figure 5 illustrates the solution of
(\ref{fSurface}) and (\ref{fg1.2}) graphically for several
temperatures at fixed doping. The solution of Eq.\ (\ref{fg1.2}) is
represented by the curves bounding the shaded regions and the
solution of Eq.\ (\ref{fSurface}) is represented by the Fermi
surface curves in each corner. The intersection of these curves
defines two simultaneous solutions in each of the four quadrants
that bound the surviving part of the Fermi surface (heavier portions
of the curves in Fig.\ 5).  In the shaded portions of Fig.\ 5 the
Fermi surface has been destroyed by the pseudogap, leaving only a
vestigial Fermi arc between the shaded regions.

The solution in Fig.\ 5 represents a derivation of Fermi-arc
structure expected in the underdoped regime above $T\tsub c$. Below
$T\tsub c$, the Fermi surface is completely destroyed by the opening
of the pairing gap except at the nodal point, since the pairing gap
has no $\gamma(\theta_{k\tsub f})$ dependence [see Eq.\
(\ref{solution1})].  We emphasize that the calculations presented in
Fig.\ 5 do not involve any new parameters as long as the PG closure
temperatures have been calculated (presented in Fig.\ 4).

Kanigel {\it et al.} \cite{Kanigel06} have reported from their ARPES
experiment arc lengths for slightly underdoped Bi2212. A direct
reading of the Fermi arc length from Fig.~5 permits us to compare it
quantitatively with the Kanigel data. Details will be published
elsewhere \cite{Gu07}.

\subsection{Complete suppression of antiferromagnetism: pure
superconducting states in underdoped compounds}

\begin{figure}
  \includegraphics[height=.23\textheight]{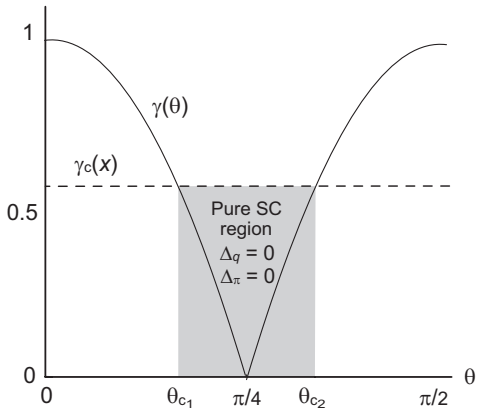}
\caption{The anisotropic factor $\gamma(\theta)$. In this figure,
$\theta_{{\scriptstyle\rm c}_{1}}=\theta\tsub c$,
$\theta_{{\scriptstyle\rm c}_{2}}=\pi/2-\theta\tsub c$, and
$\theta\tsub c$ is determined by $x_{q\theta}(\theta\tsub c)=x$.
The value of $\gamma(\theta)$ at the critical angle $\theta\tsub
c$ is denoted by $\gamma(\theta\tsub c) \equiv \gamma\tsub c(x)$,
where $\gamma\tsub c(x)$ is a monotonically increasing function of
doping $x$ that becomes equal to unity when $x=x_q$. }
\end{figure}

\begin{figure*}
  \includegraphics[height=.28\textheight]{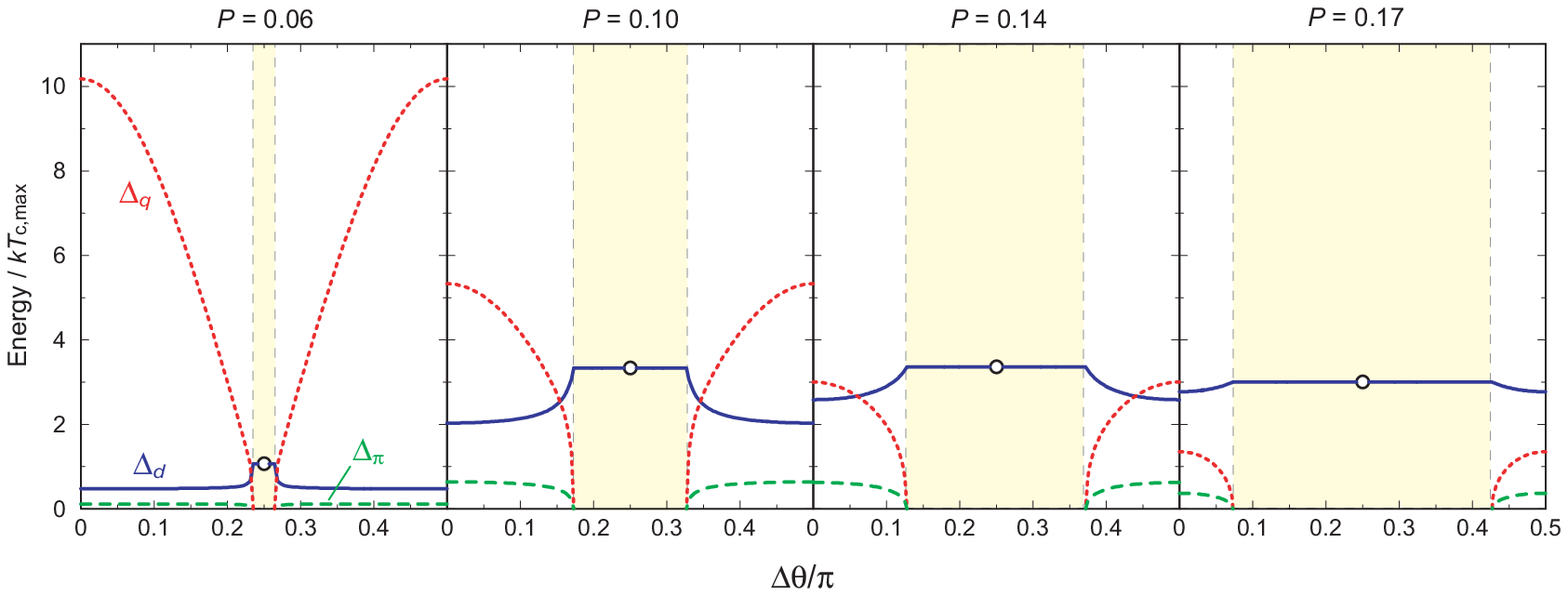}
\caption{(Color online) Dependence of the energy gaps $\Delta_d$,
$\Delta_\pi$, and $\Delta_q$ on the momentum direction $\theta$ for
several representative dopings $P$. For each doping there is a
$\theta$-window (yellow shaded region), centered at the node
$\theta=\pi/4$ (indicated by the open circle), in which
$\Delta_q=\Delta_\pi=0$, and $\Delta_d\ne 0$. This window
encompasses only a few percent of the Brillouin zone at low doping
(tending to zero at zero doping), but rapidly expands to fill the
entire Brillouin zone near the critical doping $P\simeq 0.18$ and
beyond.}
\end{figure*}

As we have discussed extensively in Ref.~\cite{Sun06}, the
antiferromagnetic correlation that plays a key role in understanding
underdoped cuprates is completely suppressed at and beyond the
critical doping point $x_q$. A pure ($d$-wave) BCS superconducting
state occurs at zero temperature in the overdoped portion of the
phase diagram.  We now show that a similar situation can also occur
in certain $k$-windows in the {\it underdoped} regime, in which AF
correlation is completely suppressed and a pure SC state emerges.

This is another interesting consequence of the SU(4)$_k$ model due
to the anisotropic factor $\gamma(\theta_{k\tsub f}) /
\overline\gamma$. The critical doping point defined in
Eq.~(\ref{xqf}), which is constant in the $k$-averaged SU(4) model
\cite{Sun06}, is now a function of momentum direction
$\theta_{k\tsub f}$ because of the anisotropic factor
$\gamma(\theta_{k\tsub f}) / \overline\gamma$. Consequently, for
each given doping $x$ there always exists a window in the momentum
azimuthal angle, $\theta\tsub c < \theta_{k\tsub f} <
(\pi/2-\theta\tsub c)$, and centering at the nodal point $\pi/4$,
within which the AF correlation vanishes and only the pairing gap
$\Delta_d$ exists. This follows because inside the window
$x_{q\theta}<x$; therefore, the solution (\ref{solution1}) is not
permitted but the solution (\ref{solution2}) is. The critical angle
$\theta\tsub c$ is determined by the condition
$x_{q\theta}(\theta\tsub c)=x$. Figure 6 illustrates the situation.

Because $\gamma(\theta_{k\tsub f}) / \overline\gamma$, and thus
$x_{q\theta}$, is a doping-dependent quantity, the above phenomenon
depends on doping. The sequential figures, plotted for four
different dopings in Fig.~7, show the behavior of the energy gaps as
functions of the momentum direction. Whenever
$$
x=x_q=\sqrt{ \frac{\chi-G_0}{\chi-G_1} },
$$
$\theta_c=0$, which means that there is no momentum space available
to $\Delta_q$ and $\Delta_\pi$, and the AF correlations and triplet
pairing states are completely suppressed. Therefore, for $x\ge x_q$
the system can only be in a pure superconducting state at zero
temperature. It can be seen that larger doping $x$ implies a smaller
critical angle $\theta\tsub c$, and thus a wider pairing window, and
that the width of the pairing window decreases rapidly toward zero
as the doping goes to zero.

In the original SU(4) model, we found that the critical doping point
defines a natural boundary (quantum phase transition) between
underdoped and overdoped regimes that have qualitatively different
wavefunctions \cite{Sun06}. We termed the underdoped superconducting
regime the AF+SC phase (antiferromagnetic superconducting phase); it
is characterized by having all gaps nonzero but is dominated by AF
and SC gaps. The present extension to the SU(4)$_k$ model reveals
the additional feature that in this AF+SC phase the gaps are {\it
highly anisotropic} in the momentum space, implying the possibility
of a pure SC window around the nodal points even in the underdoped
regime.  The proposed existence of these pure superconducting
windows may have considerable implication for the nature of the
Fermi surface at low doping, for the Nernst effect, and for the
relationship of impurities to inhomogeneities in the underdoped
region.  These deserve further investigation.

The above analysis has assumed $T=0$ for simplicity, but the basic
picture should be valid also in the case with nonzero temperature.
The formulation and solution of the gap equations described in this
paper can be extended to finite temperature using the methods
described in Ref.~\cite{Sun07}. While of considerable practical
importance, this extension does not involve conceptually new ideas
and will be deferred to a later paper.

\section{Summary}

In this paper, we have extended the SU(4) model for high-$T\tsub c$
superconductivity to include explicit momentum dependence in
observables. To do so, we have started from a general SU(4)
Hamiltonian and introduced a new approximation for the $d$-wave
formfactor in the pair operators. This leads to the new SU(4)$_k$
model, which retains explicit $k$-dependence while preserving SU(4)
symmetry. We have solved the gap equations derived from the
SU(4)$_k$ coherent states with some plausible approximations,
obtaining analytical solutions for $k$-dependent superconducting
gaps, pseudogaps, and their transition temperatures $T\tsub c$ and
$T^*$. The new SU(4)$_k$ model reduces to the original SU(4) model
for observables that are averaged over all possible $k$ directions.
Therefore we propose that the original SU(4) model describes the
averaged features and thermal properties of cuprates, while the new
SU(4)$_k$ model presented in this paper extends this description to
detailed anisotropic properties in the $k$-space. The present
results have been obtained for zero temperature but the formalism
presented here may be extended to finite temperature in a manner
similar to that extension of the $k$-averaged SU(4) model.

Because of an anisotropic factor $\gamma(\theta_{k\tsub
f})/\overline\gamma$ in the analytical gap solutions, the cuprate
phase structure in the underdoped regime becomes even richer than
that for $k$-averaged observations. We have discussed three
immediate consequences that emerge in the new SU(4)$_k$ model:
\begin{enumerate}

\item We have suggested the possibility of two distinct, measurable,
pseudogap closure temperatures: the maximum and the averaged. In the
coherent-state SU(4) theory the pseudogap could be interpreted
either as arising from competing AF and SC degrees of freedom, or
alternatively as fluctuations of pairing subject to SU(4)
constraints \cite{Sun06,Sun07}. The proposed $T^*_{\rm ave}$ and
$T^*_{\rm max}$ share this same microscopic origin, but differ from
each other by a doping-dependent factor. The temperature $T^*_{\rm
ave}$ represents PG closure temperatures that are averages over $k$,
while $T^*_{\rm max}$, which is generally higher than $T^*_{\rm
ave}$, represents the pseudogap temperature expected if one retains
explicit $k$-dependence. Experimentally, then, we predict that
$T^*_{\rm ave}$ is the pseudogap temperature that should be measured
in experiments that do not resolve $k$ explicitly, but (the
generally higher) $T^*_{\rm max}$ is the expected measured pseudogap
temperature for experiments like ARPES that resolve $k$.

\item We have provided a theoretical framework to understand ARPES
Fermi-arc data. Using two analytical equations, we have obtained
solution for the $T/T^*$ dependent Fermi-arc lengths in quantitative
agreement with existing measurements. The essence of this result is
the appearance of the new factor $\gamma(\theta)$ in the PG closure
temperature that has been derived in Eq.~(\ref{Tstark}).

\item We have predicted the existence of doping-dependent windows
in the momentum space where antiferromagnetic correlation is
completely suppressed in the underdoped regime. Without AF
competition, it is possible for pure superconducting states to
emerge in these windows. Thus, we find that pure BCS-type
superconducting states can exist, not only in conventional
superconductors or in overdoped cuprate high-$T\tsub c$
superconductors (where such behavior is well established), but in
localized islands even in underdoped cuprate superconductors. It is
of interest whether this prediction is related to the recent
observation of small pockets of well-defined fermi surface in
underdoped cuprate superconductors.

\end{enumerate}

\noindent Some of these predictions (for example, the two
temperature scales for pseudogap behavior) have the potential to
reconcile apparent discrepancies in existing data.  All make
predictions that can be tested in experiments capable of resolving
$k$-dependent behavior.

Finally, we note that the recent discovery of superconductivity in
layered iron-based transition metal oxypnictides \cite{new1} has
generated a new wave of research interest. In place of copper and
oxygen, the new compounds contain iron and arsenic, and the highest
critical temperature for them has already reached 55 kelvin
\cite{new2}. It has been demonstrated in neutron-scattering
experiments \cite{Dai08} that, like high-$T_c$ copper oxides,
superconductivity in these iron-based materials is likely competing
strongly with antiferromagnetic degrees of freedom. It will be of
considerable interest to see whether approaches like the one
presented in this paper, or other models capable of handling
multiple competing degrees of freedom in strongly-correlated systems
on an equal footing, can explain these new high temperature
superconductors and their relationship to the old ones \cite{Sun08}.
In particular, we note that what is already known about the new
iron-based superconductors suggests that $k$-dependent phenomena of
the sort described in this paper should also be observable in these
new superconductors.

%\vfill

%---------------------------------------------------------------
\bibliographystyle{unsrt}

\begin{thebibliography} {99}
\bibitem{spinNote} We assume that total spin is conserved and is zero
for the ground state.  Charge-density wave and spin-wave degrees of
freedom are ignored at the minimal level.
\end{thebibliography}

\end{document}